\documentstyle[12pt]{article}
\begin{document}
\title{\bf Interpretation of the nonextensitivity parameter
$q$ in some applications of Tsallis statistics and L\'evy
distributions}  
\author{G.Wilk$^{1}$\thanks{e-mail: wilk@fuw.edu.pl} and Z.W\l
odarczyk$^{2}$ \thanks{e-mail: wlod@pu.kielce.pl}\\[2ex] 
$^1${\it The Andrzej So\l tan Institute for Nuclear Studies}\\
    {\it Ho\.za 69; 00-689 Warsaw, Poland}\\
$^2${\it Institute of Physics, Pedagogical University}\\
    {\it  Konopnickiej 15; 25-405 Kielce, Poland}}  
\date{\today}
\maketitle

\begin{abstract}
The nonextensitivity parameter $q$ occuring in some of the
applications of Tsallis statistics (known also as index of the
corresponding L\'evy  distribution) is shown to be given, in $q>1$
case, entirely by the fluctuations of the parameters of the usual
exponential distribution.\\  

[3ex]

\end{abstract}

\newpage

There is an enormous variety of physical phenomena described most
economically (by introducing only one new parameter $q$) and 
adequately by the so called nonextensive statistics introduced
some time ago by Tsallis \cite{T}. They include all situations
characterized by long-range interactions, long-range microscopic
memories and space-time (and phase-space as well) (multi) fractal
structure of the process (cf. \cite{T} for details). The high energy
physics applications of nonextensive statistics are quite recent, but
already numerous and still growing, cf. Refs.
\cite{WW,BCM,FLUQ,UWW,RHIP,RAF,FOOT1}. All examples mentioned above
have one thing in common: the central formula employed is the
following power-like distribution:  
\begin{equation}
G_q(x)\, =\, C_q\, \left[\, 1\, -
                   \, (1\, -\, q)\, \frac{x}{\lambda}\,
                   \right]^{\frac{1}{1 - q}}, \label{eq:T}
\end{equation}                   
which is just a one parameter generalization of the Boltzmann-Gibbs
exponential formula to which it converges for $q\rightarrow 1$:
\begin{equation}
G_{q=1}\, =\, g(x)\, =\, c\cdot \exp\left[\, -\, \frac{x}{\lambda}\,
\right]                \label{eq:BG}
\end{equation}
When $G_q(x)$ is used as probability distribution (L\'evy
distribution) of the variable $x\in(0, \infty)$ (which will be the
case we are interested in here), the parameter $q$ is limited to $1\leq q
< 2$. For $q<1$, the distribution $G_q(x)$ is defined only for $x\in [0,
\lambda/(1-q)]$. For $q>1$ the upper limit comes from the
normalization condition (to unity) for $G_q(x)$ and from the
requirement of the positivity of the resulting normalisation constant
$C_q$. However, if one demands in addition that the mean value of
$G_q(x)$ is well defined, i.e., that $\langle x\rangle = \lambda
/(3-2q) < \infty$ for $x\in (0,\infty)$, then $q$ is further limited
to $1\leq q< 1.5$ only. In spite of numerous applications of the
L\'evy distribution $G_q(x)$, the interpretation of the parameter $q$
is still an open issue. In this work we shall demonstrate, on the
basis of our previous application of the L\'evy distribution to
cosmic rays \cite{WW}, that this L\'evy distribution $G_q(x)$
(\ref{eq:T}) emerges in a natural way from the fluctuations of the
parameter $1/\lambda$ of the original exponential distribution
(\ref{eq:BG}) and that the parameters of its distribution
$f(1/\lambda)$ define parameter $q$ in unique way.\\

Let us first briefly summarise the result of \cite{WW}. Analysing
experimental distributions $dN(x)/dx$ of depths $x$ of interactions
of hadrons from cosmic ray cascades in the emulsion chambers, we
have shown that the so called {\it long flying component}
(manifesting itself in aparently unnexpected nonexponential
behaviour of  $dN(x)/dx$) is just a manifestation of the L\'evy
distribution $G_q(x)$ with $q=1.3$. This result must be confronted
with our earlier analysis of the same phenomenon \cite{WWCR}. We have
demonstrated there that distributions $dN(x)/dx$ can be also 
described by the fluctuation of the corresponding cross-section
$\sigma\, =\, A\, m_N\, \frac{1}{\lambda}$ (where $A$ denotes mass
number of the target, $m_N$ is the mass of the nucleon and $\lambda$
is the corresponding mean free path). The fluctuation of this
cross-section (i.e., in effect, fluctuations of the quantity
$1/\lambda$) with relative variance  
\begin{equation}
\omega \, =\, \frac{\langle \sigma^2 \rangle\, -\, \langle
\sigma\rangle ^2}{\langle \sigma \rangle ^2} \, \ge \, 0.2 \label{eq:OMEGA}
\end{equation}
allow us to describe the non-exponentiality of the experimental data
as well as the distribution $G_{q=1.3}(x)$ mentioned above.
We therefore argue that these two numerical examples show that
fluctuations of the parameter $1/\lambda$ in the $g(x;\lambda)$
result in the L\'evy distributions $G_q(x;\lambda)$.\\

Actually the above quoted example from cosmic ray physics is not the
only one known at present in the field of high energy collisions. It
turns out \cite{BCM,UWW} that distributions of transverse momenta
$dN(p_T)/dp_T$ are best described by a slightly non-exponential
distribution $G_q(p_T)$ of the L\'evy type with $q=1.01\div1.2$
depending on situation considered. The usual exponential distribution
$dN(p_T)/dp_T = g(p_T) \sim \exp( - \sqrt{m^2+p_T^2}/kT)$ contains as
a main parameter the inverse temperature $\beta = 1/kT$ and the above
mentioned numerical results leading to $G_{q=1.01\div1.2}(p_T)$ can
again be understood as a result of a fluctuation of inverse
temperature $\beta$ in the usual exponential formula $g(p_T)$. This
point is of special interest because of recent discussions on the
dynamical possibility of temperature fluctuations in some collisions,
cf. Ref. \cite{LS,FLUCT,L}. Later on we shall use it to illustrate
our results concerning $q$.\\   

To recapitulate: we claim that (for $q>1$) the parameter $q$ is
nothing but a measure of fluctuations present in L\'evy distributions
$G_q(x)$ describing particular processes under consideration. To make
our statement more quantitative, let us analyse the influence of
fluctuations of parameter $1/\lambda$ which are present in the
exponential formula $g(x) \sim \exp(-x/\lambda)$ on the final result.
Our aim will be a deduction of the form of the function
$f(1/\lambda)$ which leads from an exponential distribution $g(x)$ to
power-like L\'evy distribution $G_q(x)$ and which describes
fluctuation about the mean value $1/\lambda_0$, i.e., such that 
\begin{equation}
G_q(x;\lambda_0)\, =\, C_q\, 
\left( 1\, +\, \frac{x}{\lambda_0}\, \frac{1}{\alpha}\right)^{-a}\, =\,
C_q\, \int^{\infty}_0\, \exp\left( - \frac{x}{\lambda}\right)\,
f\left(\frac{1}{\lambda}\right)\, d\left(\frac{1}{\lambda}\right)
\label{eq:DEF} 
\end{equation}
where for simplicity we have introduced the abbreviation $\alpha =
\frac{1}{q-1}$. 
From the representation of the Euler gamma function we have
\cite{FOOT2}
\begin{equation}
\left( 1\, +\, \frac{x}{\lambda_0}\, \frac{1}{\alpha}\right)^{-a}\, =\,
\frac{1}{\Gamma(\alpha)}\, \int^{\infty}_0\, d\xi\, 
      {\xi}^{\alpha - 1}\, \exp\left[ - \xi\, 
      \left(1\, +\, \frac{x}{\lambda_0}\,
       \frac{1}{\alpha}\right)\right] .
 \label{eq:GF}
\end{equation}
Changing variables under the integral in such a way that
$\frac{\xi}{\lambda_0}\frac{1}{\alpha}=\frac{1}{\lambda}$ one
immediately obtains eq. (\ref{eq:DEF}) with $f(1/\lambda)$ given by
the following gamma distribution 
\begin{equation}
f\left(\frac{1}{\lambda}\right)\, =\,
f_{\alpha}\left(\frac{1}{\lambda},\frac{1}{\lambda_0}\right)\, =\,
\frac{1}{\Gamma(\alpha)}\,  (\alpha\lambda_0)\,
\left(\frac{\alpha\lambda_0}{\lambda}\right)^{\alpha-1}\, \exp\left(
- \frac{\alpha\lambda_0}{\lambda}\right) \label{eq:F}
\end{equation}
with mean value
\begin{equation}
\left\langle \frac{1}{\lambda}\right\rangle \, =\,
 \frac{1}{\lambda_0} \label{eq:MEAN}
\end{equation}
and variation
\begin{equation}
\left\langle \left(\frac{1}{\lambda}\right)^2\right\rangle\, -\, 
\left\langle\frac{1}{\lambda}\right\rangle^2\, =\, 
\frac{1}{\alpha\, \lambda_0^2} . \label{eq:VAR}
\end{equation}
Notice that, with increasing $\alpha$ variance (\ref{eq:VAR})
decreases and asymptotically (for $\alpha \rightarrow \infty$, i.e,
for $q\rightarrow 1$) the gamma distribution (\ref{eq:F}) becomes
a delta function $\delta (\lambda - \lambda_0)$. The relative
variance (cf. eq.(\ref{eq:OMEGA})) for this distribution is given by 
\begin{equation}
\omega\, =\, \frac{\left\langle\left(\frac{1}{\lambda}\right)^2\right\rangle\,
 -\, \left\langle\frac{1}{\lambda}\right\rangle^2}
 {\left\langle \frac{1}{\lambda}\right\rangle^2}\, =\,
\frac{1}{\alpha}\, =\, q\, -\, 1 . \label{eq:PROOF}
\end{equation}
We see therefore that, indeed, the parameter $q$ in the L\'evy
distribution $G_q(x)$ describes the relative variance of the
parameter $1/\lambda$ present in the exponential distribution
$g(x;\lambda)$.\\ 

Some remarks on the numerical results quoted before \cite{WW,WWCR}
are in order here. Notice that the value of $q=1.3$ for cosmic ray
distribution $dN(x)/dx$ obtained in \cite{WW} leads to the relative
variance of the cross section $\omega = 0.3$ whereas in \cite{WWCR}
we have reported value $\omega' =0.2$. This discrepancy has its
origin in the fact that in numerical calculations in \cite{WWCR} we
have used a symmetric Gaussian distribution to decribe fluctuations
of the cross section, whereas the relation (\ref{eq:PROOF}) has been
obtained for fluctuations described by gamma distribution. In the
gaussian approximation we expect that  
\begin{equation}
\frac{q - 1}{q^2}\, <\, \omega'\, <\, q - 1 , \label{eq:INEQ}
\end{equation}
where lower and upper limits are obtained by normalizing the variance
of the $f(1/\lambda)$ distribution to the modial (equal to
$(2-q)/\lambda_0$) and mean (equal to $1/\lambda_0$) values,
respectively. Therefore for $q=1.3$ one should expect that $ 0.18 <
\omega' < 0.3$, which is exactly the case.\\

Let us now proceed to the above mentioned analysis of transverse
momentum distributions in heavy ion collisions, $dN(p_T)/dp_T$
\cite{FLUQ}. It is interesting to notice that the relatively small
value $q \simeq 1.015$ of the nonextensive parameter obtained there,
if interpreted in the same spirit as above, indicates that rather
large relative fluctuations of temperature, of the order of $\Delta
T/T \simeq 0.12$, exist in nuclear collisions. It could mean
therefore that we are dealing here with some fluctuations existing in
small parts of the system in respect to the whole system (according
to interpretation of \cite{L}) rather than with fluctuations of the
event-by-event type in which, for large multiplicity $N$,
fluctuations $\Delta T/T = 0.06/ \sqrt{N}$ should be negligibly small
\cite{LS}.\\ 

We shall now propose a general explanation of the meaning of the
function $f(\chi)$ describing fluctuations of some variable $\chi$.
In paticular, we shall be interested in question why, and under what
circumstances, it is the gamma distribution that describes
fluctuations. To this end let us start with the well known equation
for the variable $\chi$, which in the Langevin formulation has the
following form \cite{FP}  
\begin{equation}
\frac{d\chi}{dt}\, +\, \left[\frac{1}{\tau}\, +\, \xi(t)\right]\,
\chi\, =\, \phi\, =\, {\rm const}\, >\, 0 . \label{eq:LE}
\end{equation}
Let us concentrate for our purposes on the stochastic process which is
defined by the {\it white gaussian noise} $\xi(t)$ with ensemble mean 
\begin{equation}
\langle \xi(t) \rangle\, =\, 0 \label{eq:EM}
\end{equation}
and correlator $\langle \xi(t)\, \xi(t + \Delta t) \rangle$, which
for sufficiently fast changes is equal to 
\begin{equation}
\langle \xi(t)\, \xi(t + \Delta t) \rangle\, =\, 2\, D\,
\delta(\Delta t) .\label{eq:COR}
\end{equation}
Constants $\tau$ and $D$ define, respectively, the mean time for
changes and their variance by means of the following conditions:
\begin{equation}
\langle \chi(t)\rangle\, =\, \chi_0\, \exp\left( - \frac{t}{\tau} \right)
\quad {\rm and} \quad
\langle \chi^2(t=\infty)\rangle\, =\, \frac{1}{2}\, D\, \tau .
\label{eq:COND}
\end{equation}
Thermodynamical equilibrium is assumed here (i.e., $t >> \tau$, in
which case the influence of the initial condition $\chi_0$ vanishes
and the mean squared of $\chi$ has value corresponding to the state of
equilibrium). Making use of the Fokker-Plank equation \cite{FOOT3}
\begin{equation}
\frac{df(\chi)}{dt}\, =\, -\, \frac{\partial}{\partial \chi}K_1\,
f(\chi)\, +\, \frac{1}{2}\, \frac{\partial^2}{\partial \chi^2}K_2\,
f(\chi)  \label{eq:FPE}
\end{equation}
we get for the distribution function the following  expression
\begin{equation}
f(\chi)\, =\, \frac{c}{K_2(\chi)}\, \exp\left[\, 2\,
\int^{\chi}_0 d\chi'\, \frac{K_1(\chi')}{K_2(\chi')}\, \right]
\label{eq:EF} 
\end{equation}
where the constant $c$ is defined by the normalisation condition for
$f(\chi)$: $\int^{\infty}_0 d\chi f(\chi) = 1$. $K_1$ and $K_2$
are the intensity coefficients which for the process defined by eq.
(\ref{eq:LE}) are equal to (cf., for example, \cite{ADT}):
\begin{eqnarray}
K_1(\chi)\, &=&\, \phi\, -\, 2\, \frac{\chi}{\tau}\, +\, D\, \chi
,\nonumber\\ 
K_2(\chi)\, &=&\, 2\, D\, \chi^2 . \label{eq:KK}
\end{eqnarray}
It means therefore that as result we have the following distribution
function 
\begin{equation}
f(\chi)\, =\, \frac{1}{\Gamma(\alpha)}\, \mu\, 
 \left(\frac{\mu}{\chi}\right)^{\alpha-1}\, \exp\left( -\,
\frac{\mu}{\chi} \right) , \label{eq:FRES}
\end{equation}
which is nothing but a gamma distribution of variable $1/\chi$
depending on two parameters:  
\begin{equation}
\mu\, =\, \frac{\phi}{D} \qquad {\rm and} \qquad \alpha\, =\,
\frac{1}{\tau\, D} . \label{eq:PAR}
\end{equation}
Returning to the $q$-notation (cf. eq. (\ref{eq:DEF})) we have
therefore 
\begin{equation}
q\, =\, 1\, +\, \tau\, D , \label{eq:DEFQ}
\end{equation}
i.e., the parameter of nonextensitivity is given by the parameter $D$
describing the {\it white noise} and by the damping constant $\tau$.
This means then that the relative variance $\omega(1/\chi)$ of distribution
(\ref{eq:FRES}) is (as in eq. (\ref{eq:PROOF})) given by $\tau D$.\\

As illustration of the genesis of eq. (\ref{eq:LE}) used to derive
eq. (\ref{eq:DEFQ}), we turn once more to the fluctuations of
temperature \cite{LS,FLUCT,L} discussed before (i.e., to the
situation when $\chi = T$). Suppose that we have a thermodynamic
system, in a small (mentally separated) part of which the temperature
fluctuates with $\Delta T \sim T$. Let $\xi(t)$ describes stochastic
changes of temperature in time. If the mean temperature of the system
$\langle T\rangle = T_0$ then, as result of fluctuations in some
small selected region, the actual temperature $T'$ equals 
\begin{equation}
T'\, =\, T_0\, -\, b\, \xi(t)\, T  ,\label{eq:TTT}
\end{equation}
where the constant $b$ is defined by the actual definition of the
stochastic process under consideration, i.e., by $\xi(t)$, which is
assumed to satisfy conditions given by eqs. (\ref{eq:EM}) and
(\ref{eq:COR}). The inevitable exchange of heat between this selected
region and the rest of the system leads to the equilibration of the
temperature. The corresponding process of heat conductance is
described by the following equation \cite{LLH} 
\begin{equation}
c_p\, \rho\, \frac{\partial T}{\partial t}\, -\, a\, (T'\, -\, T)\,
=\, 0 , \label{eq:HC}
\end{equation}
where $c_p,~\rho$ and $a$ are, respectively, the specific heat,
density and the coefficient of the external conductance. Using $T'$
as defined in (\ref{eq:TTT}) we finally get the linear differential
equation (\ref{eq:LE}) for the temperature $T$ with coefficients:
$\tau = \frac{c_p\rho}{a}$, $\phi = \frac{a}{c_p\rho}T_0 = T_0/\tau$
and $b=\tau$:
\begin{equation}
\frac{\partial T}{\partial t}\, +\, \left[\, \frac{a}{c_p\, \rho}\,
+\, \frac{a}{c_p\, \rho}\, b\, \xi(t)\, \right]\, T\, =
\, \frac{a}{c_p\, \rho}\, T_0 . \label{eq:RESF}
\end{equation}
This result demonstrates clearly that one can think of a deep
physical interpretation of the parameter $q$ of the corresponding
L\'evy distribution describing the distributions of the transverse
momenta mentioned before. In this respect our work differs from works
in which $G_q(x)$ is shown to be connected with $G_{q=1}(x) = g(x)$
by the so called Hilhorst integral formula (the trace of which
is our eq. (\ref{eq:GF})) \cite{FOOT2,C} but without discussing the
physical context of the problem. Our original motivation was to
understand the apparent success of Tsallis statistics (i.e., the
situations in which $q>1$) in the realm of high energy collisions.\\

To summarise: if fluctuations of the variable $\chi$ can be described
in terms of the Langevin formulation, their distribution function
$f(1/\chi)$ satisfies the Fokker-Plank equation and is therefore
given by the Gamma distribution in the variable $1/\chi$. Such
fluctuations of the parameter $1/\chi$ in the exponential formula of
physical interest, $g(x/\chi)$, lead immediately to a L\'evy
distribution $G_{q>1}(x/\chi)$ with $q$ parameter given by the
relative variance of the fluctuations described by $f(1/\chi)$. 
It should be stressed that in this way we address the interpretation
of only very limited cases of applications of Tsallis statistics.
They belong to the category in which the power laws physically appear
as a consequence of some continuous spectra within appropriate
integrals. It does not touche, however, a really {\it hard} case of
applicability of Tsallis statistics, namely when {\it zero} Lyapunov
exponents are involved \cite{FOOT4}. Nevertheless, this allows us to
interpret some nuclear collisions data in terms of fluctuations of
the inverse temperature, providing thus an important hint to the
origin of some systematics in the data, understanding of which is
crucial in the search for the new state of matter: the Quark Gluon
Plasma \cite{FLUQ,FLUCT}.\\ 

Acknowledgement: We are grateful to Prof. St. Mr\'owczy\'nski for 
fruitful discussions and comments.

\end{document}